\documentclass[twocolumn]{article}
\usepackage{microtype}
\usepackage[centertags]{amsmath}
\usepackage{amssymb}
\usepackage{url}
\usepackage{graphicx}
\graphicspath{{Figures/}}
\usepackage{enumitem}
\usepackage[table]{xcolor}
\usepackage{booktabs,tabularx,array}
\usepackage{verbatim} 
\usepackage[export]{adjustbox}
\usepackage{float}
\usepackage{caption}
\usepackage{subcaption}
\usepackage{mathtools}
\usepackage{amsthm}
\usepackage{cancel}
\usepackage{hyperref}
\usepackage{pdfpages}
\usepackage{hyperref}
\usepackage{graphicx}
\graphicspath{{Figures/}}
\usepackage[export]{adjustbox}
\usepackage{float}
\usepackage{caption}
\usepackage{subcaption}
\usepackage{mathtools}
\usepackage{amsthm}
\usepackage{cancel}
\usepackage{hyperref}
\usepackage{pdfpages}
\usepackage{bm}

\def\BibTeX{{\rm B\kern-.05em{\sc i\kern-.025em b}\kern-.08em
		T\kern-.1667em\lower.7ex\hbox{E}\kern-.125emX}}

\def\lef[{\left[\begin{array}}
\def\rig]{\end{array}\right]}

\usepackage{psfrag}

\def\begequ{\begin{equation}}
\def\endequ{\end{equation}}

\def\begite{\begin{itemize}}
\def\endite{\end{itemize}}
\def\begarr{\begin{array}}
\def\endarr{\end{array}}
\def\begequarr{\begin{eqnarray}}
\def\endequarr{\end{eqnarray}}

\usepackage{multicol}

\def\L2{{\cal L}_2}
\def\L2e{{\cal L}_{2e}}

\def\begsubequ{\begin{subequations}}
\def\endsubequ{\end{subequations}}
\def\begequarr{\begin{eqnarray}}
\def\endequarr{\end{eqnarray}}
\def\begequarrs{\begin{eqnarray*}}
\def\endequarrs{\end{eqnarray*}}
\def\begarr{\begin{array}}
\def\endarr{\end{array}}
\def\begequ{\begin{equation}}
\def\endequ{\end{equation}}

\def\begdes{\begin{description}}
\def\enddes{\end{description}}
\def\begenu{\begin{enumerate}}
\def\begite{\begin{itemize}}
\def\endite{\end{itemize}}
\def\endenu{\end{enumerate}}

\def\lef[{\left[\begin{array}}
\def\rig]{\end{array}\right]}

\def\begcen{\begin{center}}
\def\endcen{\end{center}}
\def\begrem{\begin{remark}\rm}
\def\endrem{\end{remark}}
\def\begassums{\begin{assumption*}}
\def\endassums{\end{assumption*}}
\def\begassu{\begin{ass}}
\def\endassu{\end{ass}}
\def\beglem{\begin{lemma}}
\def\endlem{\end{lemma}}
\def\begcor{\begin{corollary}}
\def\endcor{\end{corollary}}
\def\begfac{\begin{fact}}
\def\endfac{\end{fact}}
\def\begass{\begin{assumption}}
\def\endass{\end{assumption}}

\usepackage{color}

\graphicspath{{Images}}
\hypersetup{
    colorlinks=true,
    linkcolor=blue,
    filecolor=blue,
    urlcolor=blue,
    citecolor=blue,
    }

\title{A Response to the Reply Presented in ``A Fair Comparison of Sliding-Mode and
Immersion-and-Invariance Observers''} 

\author{Luis Cervantes-Pérez*\textsuperscript{1}, Víctor Santibáñez\textsuperscript{1}, Jesús Sandoval\textsuperscript{2},\\
 Romeo Ortega\textsuperscript{3}, and Jose Guadalupe Romero\textsuperscript{3}\\
  \small 1 Tecnológico Nacional de México / I. T. de La Laguna,\\
  \small Torreón, Coahuila, México \\
 \small 2 Tecnológico Nacional de México / I. T. de La Paz,\\
 \small La Paz, Baja California Sur, México \\
 \small 3 Departamento de Ingeniería Eléctrica y Electrónica / \\
 \small Instituto Tecnológico Autónomo de México, CDMX, Ciudad de México, México
 }
\usepackage{titlesec}
\date{}

\titleformat{\paragraph}
{\normalfont\normalsize\bfseries}{\theparagraph}{1em}{}
\titlespacing*{\paragraph}
{0pt}{3.25ex plus 1ex minus .2ex}{1.5ex plus .2ex}

\makeatletter
\AtBeginDocument{\let\@cite@ofmt\@firstofone}
\makeatother
\begin{document}
\twocolumn[
  \begin{@twocolumnfalse}
    \maketitle
    \begin{abstract}
In response to the reply presented in \cite{FairComparison}, this technical report extends the experimental evaluation of position-feedback velocity observers and adaptive tracking controllers for a pendular system with nonlinear friction reported in \cite{CervantesPerez2026Experimental}. The immersion-and-invariance (I$\&$I) adaptive control scheme proposed in \cite{romero2025adaptive}, whose gains are tuned according to the guidelines presented in \cite{ReflectionsSM}, is compared with a control scheme employing the adaptive modified super-twisting observer of \cite{SM1}, whose observer gains are selected according to the \emph{guidelines} proposed in \cite{FairComparison}. Although these ``tuning guidelines'' attenuate chattering in certain scenarios, the experimental results reported here show that they do not eliminate it in general.
    \end{abstract}
  \end{@twocolumnfalse}
]

\section{Experimental platform and scope}\label{secc:ExperimentalEvaluation}
\begin{figure}[h!]
  \centering
  \includegraphics[scale=0.2]{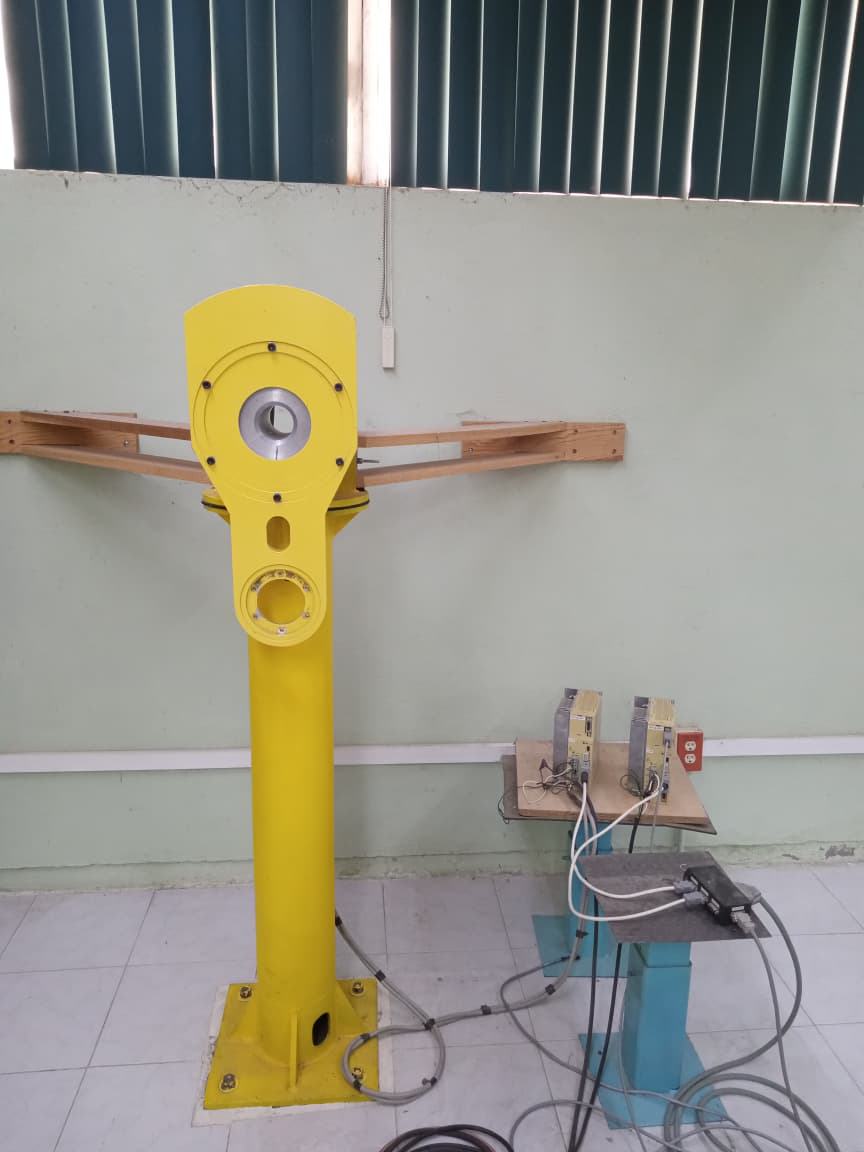}
  \caption{Pendular device.}
    \label{fig:robot}
\end{figure}

\begin{figure}[h!]
  \centering
  \includegraphics[width=0.3\textwidth]{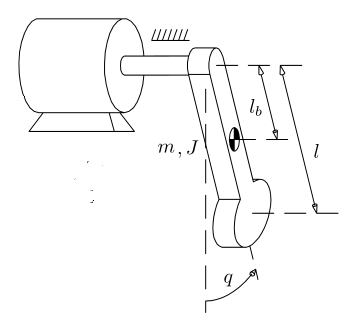}
  \caption{Sketch of the pendular device.}
    \label{fig:sketch}
\end{figure}

\begin{table*}
	\caption{Physical parameters of the pendular device.}
	\label{tab:parrob}
	\begin{center}
		\begin{tabular}{cccc} \hline
			Description  & Notation & Value & Units \\ \hline
			Length of Link  & $l$ & 0.313 & $\mathrm{m}$ \\
			Distance to the center of mass  & $l_{b}$ & 0.0641 & $\mathrm{m}$  \\
		Mass of Link  & $m$ & 22.4466 & $\mathrm{K_g}$ \\
		   Inertia relative to center of mass  & $J$ & 0.7013  & $\rm Kg\ m^2$ \\
          Viscous friction coefficient & $\theta_1$ & $5.317$ &  $\mathrm{Nm \ s/rad}$ \\
          Coulomb friction coefficient & $\theta_2$ & $11.6403$ &  $\mathrm{Nm }$ \\
			Gravity acceleration constant & $g$ & 9.81 & $\frac{{\rm m}}{{\rm s^2}}$ \\ \hline
		\end{tabular}
	\end{center}
\end{table*}
This section introduces the experimental platform used to conduct the comparative study and outlines the scope of the reported tests. The testbed consists of a direct-drive pendular device with a single vertical link, designed and built at the La Laguna Technological Institute, Mexico. A picture of the pendular device is shown in Figure~\ref{fig:robot}, whereas its schematic diagram is presented in Figure~\ref{fig:sketch}, where $q$ denotes the angular position of the actuated joint.

\par The joint is actuated by a direct-drive motor manufactured by Yokogawa Electric Corporation, model \textit{DM1A200G}. The motor operates in torque mode, providing a maximum torque of $\pm200~\mathrm{[Nm]}$. It is also equipped with a relative encoder offering a resolution of $1{,}024{,}000$ pulses per revolution. The proposed controller was implemented digitally in MATLAB using the Simulink Desktop Real-Time environment, with a sampling period of $1~\mathrm{[ms]}$. The experiments were conducted on a PC running Windows~10, equipped with an \textit{Intel Core i5-14600K} CPU, $32~\mathrm{GB}$ of RAM, an RTX~3060 GPU, and a $512~\mathrm{GB}$ M.2 SSD. Communication between the PC and the robot was established through a Sensoray~626 data acquisition card (DAQ) installed in the PCI port of the motherboard. The DAQ interfaced with the motor driver (PARKER COMPUMOTOR model DMG3-1200A) by sending analog voltage signals corresponding to the control torques. In turn, the driver transmitted pulse trains replicating the motor encoder signals to the DAQ digital inputs. The physical parameters of the link are listed in Table~\ref{tab:parrob}, as originally reported in \cite{Gamez,Gandarilla}. The table summarizes the meaning of the parameters and their numerical values, where $l$ denotes link length, and `Dist.', `Inert.', and `com' stand for distance, inertia, and center of mass, respectively.
\section{Position-feedback adaptive velocity observers}
\label{sec:Observers}

In this section, we briefly review the two position-feedback velocity-observer structures: the I$\&$I adaptive observer proposed in \cite{romero2025adaptive} and the adaptive sliding-mode (SM) observer presented in \cite{SM1}. We present only the information relevant to the experimental evaluation. Proofs and additional details are omitted for brevity and can be found in the corresponding references.

\subsection{Pendular-device model}

The dynamics of the pendular device are described by
\begin{equation}
\label{eqn:OpenLoopStateModel}
\begin{aligned}
\dot{x}_1
&=x_2,\\
\dot{x}_2
&=
\frac{u}{J}
-\frac{K_L}{J}\sin(x_1)
-\frac{\theta_1}{J}x_2
-\frac{\theta_2}{J}\tanh(\vartheta x_2),
\end{aligned}
\end{equation}
where $x_1=q$ and $x_2=\dot q$, $u$ is the applied torque, $J$ is the moment of inertia, $\theta_1$ and $\theta_2$ are the viscous- and Coulomb-friction coefficients, respectively, and $K_L=ml_bg$.
\par Given the estimated state vector $\hat{\bm{x}}(t)\in\mathbb{R}^2$, we define the state observation error as $\tilde{\bm{x}}(t):=\hat{\bm{x}}(t)-\bm{x}(t)$.

\subsection{\texorpdfstring{I$\&$I}{I\&I} adaptive observer}

The I$\&$I adaptive observer of \cite{romero2025adaptive} is given by
\begin{align}\label{eqn:adaptive1}
\dot{x}_{2I}&=\frac{1}{J}u-\frac{ml_bg}{J}\sin(x_1)-\frac{1}{J}(\hat{\theta}_1+k_1)\hat{x}_2\\
&-\frac{1}{J}\hat{\theta}_2\tanh(\vartheta\hat{x}_2)\\
\dot{\theta}_{1I}&=\gamma_1  \frac{\vartheta}{k_1}\hat{x}_2\Big(\frac{k_1}{J}\hat{x}_2+\dot{x}_{2I}\Big)\\
\dot{\theta}_{2I}&=\gamma_2  \frac{\vartheta}{k_1}\tanh(\vartheta \hat{x}_2) \Big(\frac{k_1}{J}\hat{x}_2+\dot{x}_{2I}\Big),\\
\hat{x}_2&=x_{2I}+\frac{k_1}{J}x_1\\
\hat{\theta}_1&=\theta_{1I}-\gamma_1 \frac{\vartheta}{2k_1}\hat{x}_2^2\\
\label{eqn:adaptive2}
\hat{\theta}_2&=\theta_{2I}-\gamma_2 \frac{1}{k_1}\ln\big(\cosh(\vartheta \hat{x}_2)\big)
\end{align}
where $k_1$, $\gamma_1$, and $\gamma_2$ are tuning parameters.

\subsection{Adaptive SM observer}

The adaptive SM observer of \cite{SM1} is expressed as
\begin{align}
\label{eqn:SMObserverA}
\dot{\hat{x}}_1(t)&=\hat{x}_2+\alpha_2\sqrt{|\tilde{x}_1|}\mathrm{sign}(\tilde{x}_1)\\
\nonumber
\dot{\hat{x}}_2(t)&=\frac{1}{J}u-\frac{ml_bg}{J}\sin(x_1)-\frac{\bar{\theta}_1}{J}\hat{x}_2-\frac{\bar{\theta}_2}{J}\tanh(\vartheta \hat{x}_2)\\
\label{eqn:SMObserverAA}
&+\alpha_1\mathrm{sign}(\tilde{x}_1)\\
\label{eqn:SMObserverAAA}
\dot{\hat{\Delta}}_{\theta}&=\Gamma\boldsymbol\varphi[-\boldsymbol\varphi^\top \hat{\Delta}_\theta +\alpha_1\mathrm{sign}(\tilde{x}_1)]\\
\label{eqn:SMObserverB}
\dot{\Gamma}&=-\Gamma\boldsymbol\varphi\boldsymbol\varphi^\top \Gamma,
\end{align}
where $\bar{\boldsymbol{\theta}}\in\mathbb{R}^2$ denotes a fixed \emph{a priori} estimate of the normalized parameter vector $\frac{1}{J}\boldsymbol{\theta}$, and $\alpha_1$ and $\alpha_2$ are positive tuning gains. The matrix $\Gamma\in\mathbb{R}^{2\times 2}$ plays a role analogous to that of the covariance matrix in the classical recursive least-squares identification method. The regressor vector $\boldsymbol{\varphi}\in\mathbb{R}^2$ is defined as
\begin{equation*}
\boldsymbol\varphi:=-
\begin{bmatrix}
 \hat{x}_2 \\ \tanh(\vartheta\hat{x}_2)
 \end{bmatrix}.
\end{equation*}
Defining $\Delta_{\theta}:=\frac{1}{J}\boldsymbol{\theta}-\bar{\boldsymbol{\theta}}$, the quantity $\hat{\Delta}_{\theta}$ denotes an estimate of the mismatch between the normalized parameter vector $\frac{1}{J}\boldsymbol{\theta}$ and its fixed \emph{a priori} estimate $\bar{\boldsymbol{\theta}}$. Consequently, an estimate of the parameter vector $\boldsymbol{\theta}$ is given by
\begin{equation*}
\hat{\boldsymbol\theta}:=J\bigg[\hat{\Delta}_{\theta}+\bar{\boldsymbol\theta}\bigg].
\end{equation*}
\subsubsection{SM observer tuning procedure}
According to \cite{FairComparison}, the SM observer gains can be selected using the prescription
\begin{equation}
\label{eqn:SMTuningCurrentTest}
\alpha_2=1.5\sqrt{L},
\qquad
\alpha_1=1.1L,
\qquad
L>\sup_{t\geq0}|\dot{\delta}(t)|.
\end{equation}
However, \cite{FairComparison} does not explicitly define the signal $\delta$, identify the term that must be interpreted as $\dot{\delta}$ for the pendular device, or provide a procedure for determining its upper bound. Verifying this condition is not straightforward because the required bound cannot be determined \emph{a priori} from the information provided. To clarify the relevance of this condition, we next derive an operational expression for $\dot{\delta}$ from the observer-error dynamics and discuss the circular dependence that arises when $L$ is inferred from trajectories generated by observer gains that themselves depend on $L$. To assign a precise meaning to $\dot{\delta}$ for the system considered here, we express the observer-error dynamics in the canonical super-twisting (ST) form used in the analyses of \cite{Levant1998,MorenoOsorio2008}.
Subtracting the plant dynamics in \eqref{eqn:OpenLoopStateModel} from the SM observer dynamics \eqref{eqn:SMObserverA}-\eqref{eqn:SMObserverAA} yields
\begin{align}
\dot{\tilde{x}}_1
&=
\tilde{x}_2-\alpha_2|\tilde{x}_1|^{1/2}
\operatorname{sign}(\tilde{x}_1),
\label{eqn:SMErrorFirst}\\
\dot{\tilde{x}}_2
&=
-\alpha_1\operatorname{sign}(\tilde{x}_1)+\varrho_2(t),
\label{eqn:SMErrorSecond}
\end{align}
where the perturbation term, expressed directly in terms of the actual and \emph{a priori} friction parameters, is
\begin{equation}
\label{eqn:SMResidualDefinition}
\begin{split}
\varrho_2(t)
:=&
\frac{\theta_1}{J}x_2(t)
+\frac{\theta_2}{J}\tanh\big(\vartheta x_2(t)\big)
-\frac{\bar{\theta}_1}{J}\hat{x}_2(t)
-\frac{\bar{\theta}_2}{J}\\
&\tanh\big(\vartheta\hat{x}_2(t)\big).
\end{split}
\end{equation}
Equations~\eqref{eqn:SMErrorFirst}--\eqref{eqn:SMErrorSecond} have the
canonical super-twisting structure in which the discontinuous injection must
dominate the perturbation entering the second error equation. Therefore, and
only for the purpose of establishing a correspondence with the notation used
in \cite{FairComparison}, we make the operational identification
$\dot{\delta}(t)\equiv\varrho_2(t)$. This expression is not postulated arbitrarily. It is precisely the term that remains in the second observer-error equation after separating the super-twisting injection. At the same time, it must be emphasized that \eqref{eqn:SMResidualDefinition} is our model-based reconstruction of the quantity that plays the role of $\dot{\delta}$; it is not a definition provided in \cite{FairComparison}. In particular, no independent signal $\delta$ is measured and subsequently differentiated. Neither the applied torque $u$ nor the gravitational term appears explicitly in \eqref{eqn:SMResidualDefinition}. Both terms cancel when the plant acceleration is subtracted from the observer dynamics because the plant and the observer use the same torque signal and gravitational model. Nevertheless, the input torque affects $\dot{\delta}$ indirectly because it modifies the trajectories of $x_2$ and $\hat{x}_2$. Consequently, changes in the input, reference trajectory, operating point, initial conditions, or external perturbations can alter the magnitude of the effective perturbation appearing in \eqref{eqn:SMErrorSecond}. Therefore, assuming that a known uniform upper bound for \eqref{eqn:SMResidualDefinition} exists and can be determined \emph{a priori} constitutes a strong and nontrivial requirement. How such a bound can be obtained or verified for the experimental system remains unclear.
 Furthermore, expression~\eqref{eqn:SMResidualDefinition} represents
only the mismatch described by the adopted nominal model. Unmodeled actuator
dynamics, measurement noise, external disturbances, parameter variations, and
discretization effects may introduce additional terms into
\eqref{eqn:SMErrorSecond}. Therefore, the experimentally reconstructed
$\dot{\delta}$ should be interpreted as a model-based diagnostic signal and
not as an independently established global bound on every perturbation acting
on the physical system.
\par In \cite{FairComparison} (see equation~(15) therein), the following so-called \emph{$L$-free relationship} is obtained by dividing $\alpha_2$ by $\sqrt{\alpha_1}$ using the tuning prescription in \eqref{eqn:SMTuningCurrentTest}, thereby eliminating $L$ and yielding
\begin{equation}
\label{eqn:LFreeRelationshipCurrent}
\frac{\alpha_2}{\sqrt{\alpha_1}}
= \frac{1.5\sqrt{L}}{\sqrt{1.1L}}
=
\frac{1.5}{\sqrt{1.1}}
\approx 1.430.
\end{equation}
However, this $L$-free relationship fixes only the ratio between the observer gains. It does not determine their absolute magnitudes and, more importantly, it does not eliminate the hypothesis $L>\sup_{t\geq0}|\dot{\delta}(t)|$ from the stability analysis on which the gain selection is based. Thus, a gain pair may satisfy \eqref{eqn:LFreeRelationshipCurrent} while the associated value of $L$ fails to dominate the perturbation generated during a particular experiment. The practical difficulty is that the required value of $L$ cannot be obtained from \eqref{eqn:LFreeRelationshipCurrent}. Moreover, under the operational definition \eqref{eqn:SMResidualDefinition}, $\dot{\delta}$ depends on $\hat{x}_2$, whose trajectory depends, in turn, on the gains $\alpha_1$ and $\alpha_2$ determined from $L$. Therefore, if $L$ is selected by first running an experiment, measuring the largest reconstructed value of $|\dot{\delta}|$, and subsequently retuning the observer, the resulting procedure is experiment-dependent and practically circular. This observation does not imply that a conditional stability proof based on a known perturbation bound is mathematically circular. Such assumptions are standard when an independent uniform bound is available beforehand. The difficulty in the present case is that \cite{FairComparison} does not derive such a bound from known plant parameters, specify an admissible set of initial conditions and inputs over which the bound must hold, or provide an invariant-set argument establishing the boundedness of the states involved in \eqref{eqn:SMResidualDefinition}. Consequently, the proposed relationship does not, by itself, constitute a complete, reproducible, and operating-condition-independent tuning procedure. One possible conservative approach would be to select $L$ as large as the practical implementation and actuator limitations permit. However, according to \eqref{eqn:SMTuningCurrentTest}, increasing $L$ also increases both observer gains: $\alpha_1$ grows linearly with $L$, whereas $\alpha_2$ grows with its square root. As shown in the experimental results presented next, this increase may intensify the chattering effects and degrade the system response, even when the gain selection satisfies \eqref{eqn:SMTuningCurrentTest}.

\begin{figure*}[ht!]
\centering
\includegraphics[width=\textwidth]{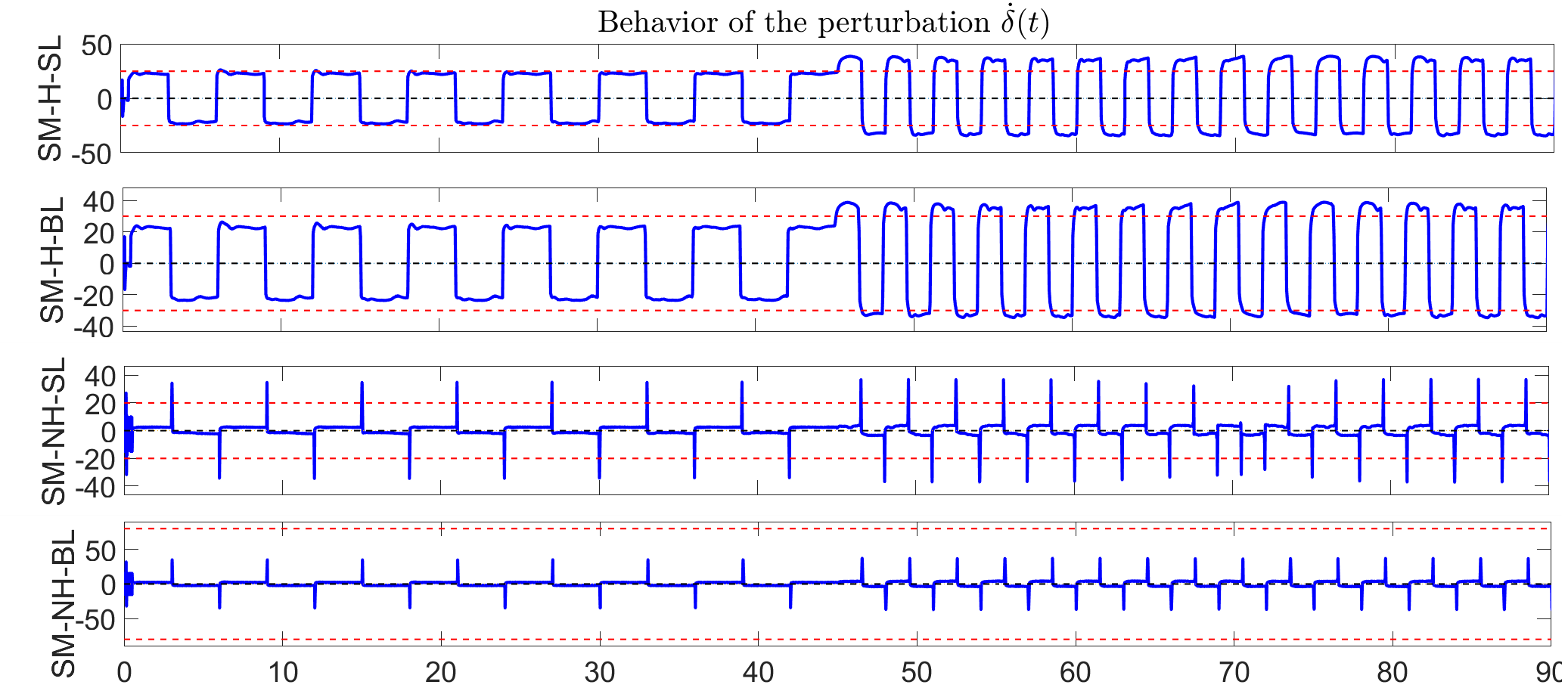}
\caption{Behavior of the reconstructed perturbation $\dot{\delta}(t)$ for the four SM configurations using different values of $L$.}
\label{fig:F1}
\end{figure*}
\section{Experimental results: open-loop test}
\label{sec:ExperimentalProtocol}
\begin{figure*}[ht!]
\centering
\includegraphics[width=\textwidth]{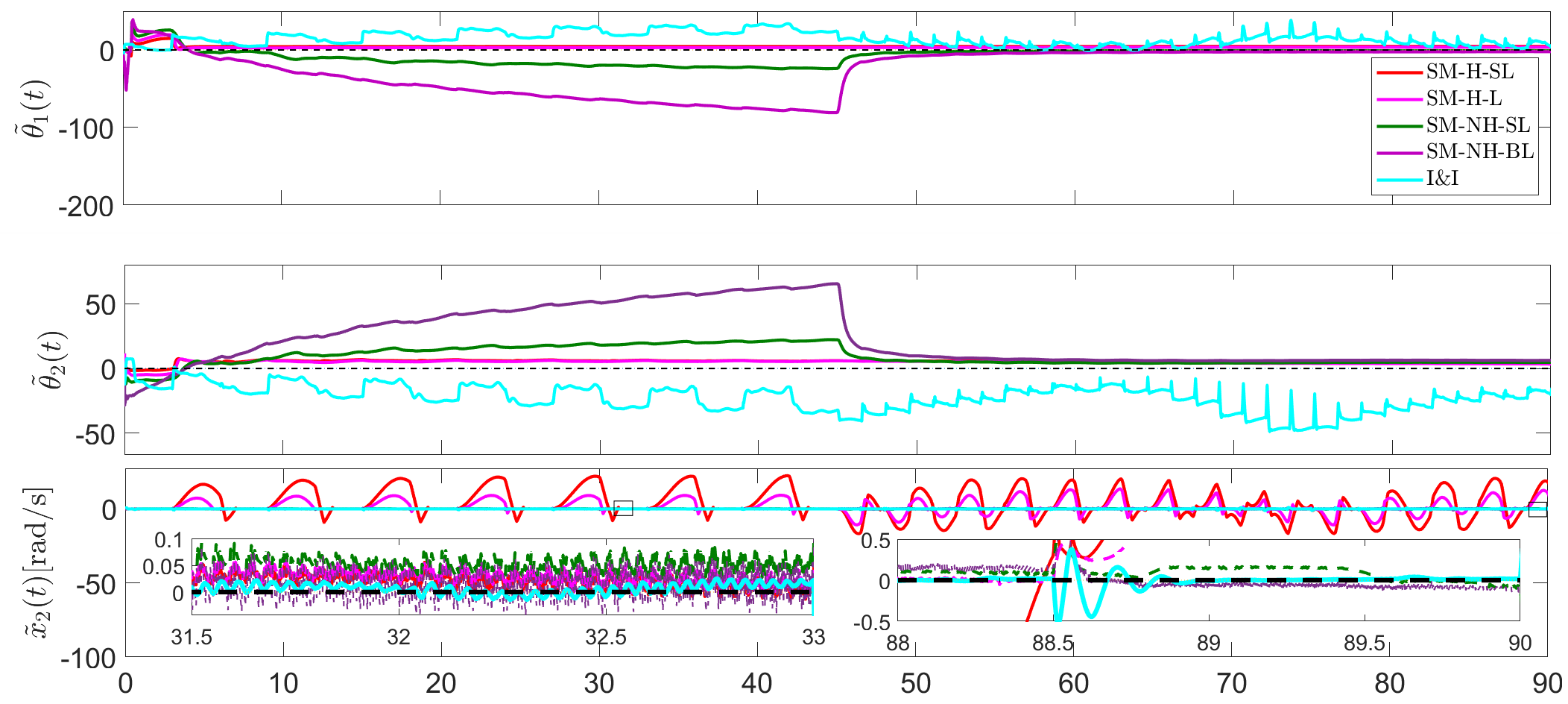}
\caption{Velocity- and friction-parameter-estimation errors for all observer configurations in the open-loop test.}
\label{fig:F3}
\end{figure*}

In this section, we present the experimental results of an open-loop test conducted to estimate the velocity and friction parameters. The experiment compares the I$\&$I observer with four configurations of the same adaptive SM observer. These configurations differ in two factors: the normalized friction coefficients specified \emph{a priori} and the value assigned to $L$ in \eqref{eqn:SMTuningCurrentTest}. To present these results, it is necessary to introduce the following notation. Following the terminology used in \cite{FairComparison}, the suffix $H$ denotes a \emph{handicapped} configuration, in which the friction parameters are considered completely unknown \emph{a priori}. For this configuration, $\frac{\bar{\theta}_1}{J}=\frac{\bar{\theta}_2}{J}=0.01$. These values represent a case in which essentially no accurate friction information is supplied to the SM observer. Conversely, the suffix $NH$ denotes the \emph{non-handicapped} configuration, for which $\frac{\bar{\theta}_1}{J}=7$ and $\frac{\bar{\theta}_2}{J}=15$. These values are close to, although not exactly equal to, the identified normalized coefficients $\frac{\bar{\theta}_1}{J}=7.5816$ and $\frac{\bar{\theta}_2}{J}=16.5989$. The purpose of considering both configurations is to evaluate the sensitivity of the observer performance to the friction information supplied \emph{a priori}. Finally, the suffixes $SL$ and $BL$ denote the smaller and larger selected values of $L$, respectively. Table~\ref{tab} summarizes the values of $L$, the resulting observer gains, and the normalized friction coefficients supplied \emph{a priori} for the four SM configurations. The gains $\alpha_1$ and $\alpha_2$ are computed according to \eqref{eqn:SMTuningCurrentTest}.
\begin{table}[ht!]
\centering
\caption{Tuning parameters for the SM configurations.}
\label{tab}
\begin{tabular}{lccccc}
\toprule
Configuration
& $L$
& $\alpha_1$
& $\alpha_2$
& $\frac{\bar{\theta}_1}{J}$
& $\frac{\bar{\theta}_2}{J}$\\
\midrule
SM-H-SL & $25$ & $27.50$ & $7.50$ & $0.01$ & $0.01$\\
SM-H-BL & $30$ & $33.00$ & $8.22$ & $0.01$ & $0.01$\\
SM-NH-SL & $20$ & $22.00$ & $6.71$ & $7$ & $15$\\
SM-NH-BL & $80$ & $88.00$ & $13.42$ & $7$ & $15$\\
\bottomrule
\end{tabular}
\end{table}
For the I$\&$I observer, we set $k=1$, $\gamma_1=1$, and $\gamma_2=1$. For all observer configurations, $\vartheta$ is set to $200$. 
\begin{figure*}[ht!]
\centering
\includegraphics[width=\textwidth]{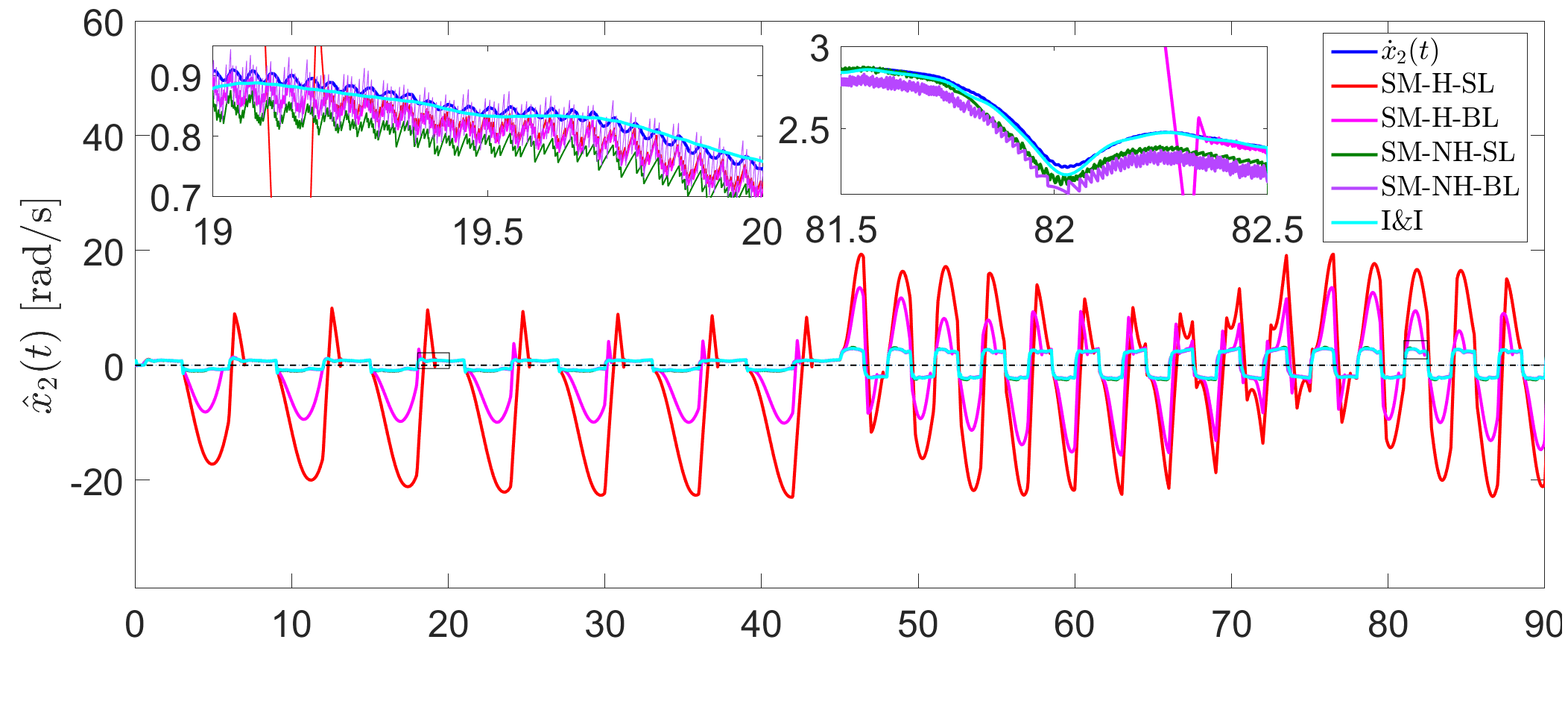}
\caption{Time behavior of the estimated velocity $\hat{x}_2(t)$ against the actual velocity }
\label{fig:F2}
\end{figure*}
All configurations are evaluated simultaneously using the same measured position and applied torque. The applied torque input is given by
\begin{equation}
\label{eqn:OpenLoopPiecewiseInput}
u(t)=
\begin{cases}
14 \operatorname{sign}\left[ \sin\left(\dfrac{\pi}{3}t\right) \right], & 0\leq t<45, \\
21 \operatorname{sign}\left[\sin\left( \dfrac{2\pi}{3} (t-45) \right) \right], & t\geq 45,
\end{cases}
\end{equation}
where $u$ is expressed in $\mathrm{[Nm]}$ and $t$ in seconds. For each SM configuration, the perturbation term $\dot{\delta}$ is reconstructed according to \eqref{eqn:SMResidualDefinition}. Figures~\ref{fig:F1}--\ref{fig:F3} show the experimental results. Figure~\ref{fig:F1} shows the behavior of the reconstructed perturbation $\dot{\delta}(t)$ for all SM observer configurations. In each case, the red dotted line represents the corresponding value of $L$, selected \emph{a priori} and regarded as a known bound in \eqref{eqn:SMTuningCurrentTest}. It should first be noted that $L$ has an intrinsic relationship with the observer-state trajectories. Indeed, $L$ determines the observer gains through \eqref{eqn:SMTuningCurrentTest}; these gains affect the trajectory of $\hat{x}_2$, which, in turn, appears explicitly in the reconstruction of $\dot{\delta}$. Likewise, comparison of the handicapped and non-handicapped configurations shows that the friction information supplied \emph{a priori} also influences the behavior of the reconstructed perturbation. The second interval of the applied torque deliberately changes both the amplitude and frequency of the input, thereby modifying the system and observer trajectories. The purpose of this change is to evaluate whether a value of $L$ that is adequate during the first operating interval remains valid under a different excitation condition.
 Therefore, this part of the experiment provides an experimental assessment of the robustness of the selected bound with respect to changes in the operating conditions. If $|\dot{\delta}(t)|$ exceeds $L$ at any instant, then the sufficient condition in \eqref{eqn:SMTuningCurrentTest} is violated during that experimental run. Conversely, if $|\dot{\delta}(t)|<L$ throughout the experiment, this only shows that the selected value of $L$ dominates the reconstructed perturbation for that particular input, initial condition, parameter set, and observation interval. It does not establish that $L>\sup_{t\geq0}|\dot{\delta}(t)|$ under other operating conditions. This distinction is central to the experiment. A value of $L$ adjusted from a particular experimental record may produce satisfactory estimation performance for that record. Nevertheless, a change in the input amplitude or frequency, a different reference trajectory in closed loop, a different initial condition, an external perturbation, or an unmodeled actuator effect may alter $\dot{\delta}$ and cause the previously selected value of $L$ to cease satisfying the required inequality. Therefore, a value calibrated from a single experiment cannot be transferred to arbitrary operating conditions without re-verifying the bound. This result should not be interpreted as claiming that every change in the operating conditions necessarily invalidates the same value of $L$. A sufficiently large uniform bound could remain valid over a prescribed family of trajectories, initial conditions, and perturbations. However, such a bound would have to be derived independently over a clearly specified operating set. Neither such an operating set nor a procedure for deriving the corresponding uniform bound is provided in \cite{FairComparison}.

Figure~\ref{fig:F2} shows the estimated velocity $\hat{x}_2(t)$ obtained with each observer configuration, including the I$\&$I observer, together with the actual velocity $x_2(t)$. The \emph{handicapped} SM configurations clearly exhibit degraded estimation performance, demonstrating the sensitivity of the observer response to the friction-parameter values supplied \emph{a priori}. For the \emph{non-handicapped} configurations, the velocity estimates follow the actual velocity more closely; however, chattering is clearly visible and becomes more pronounced as $L$ is increased. In contrast, the I$\&$I observer provides the best overall estimation performance, including after the change in the applied torque, while maintaining a smooth and continuous estimate without visible chattering.

Finally, Figure~\ref{fig:F3} shows the velocity-estimation error and the friction-parameter-estimation errors. The parameter-estimation errors do not converge to zero for any of the observer configurations, indicating that the applied excitation is not sufficiently rich to ensure parameter convergence. Regarding the velocity-estimation error, visible chattering appears in nearly all the SM configurations, in clear contrast to the smooth error signal produced by the I$\&$I observer.

\section{Experimental results: trajectory tracking test}
In this section, we describe the trajectory-tracking configuration and present the corresponding experimental results obtained using the I$\&$I and adaptive SM observers. In both cases, the control law is based on the \emph{certainty-equivalence principle} and is given by
\begin{equation}
\begin{split}
u=& \hat{\theta}_1\hat{x}_2
+\hat{\theta}_2\tanh(\vartheta\hat{x}_2)
+m l_b g\sin(x_1)
+J[\ddot{x}_{1d}\\
&-k_pe_1-k_v\left(\hat{x}_2-\dot{x}_{1d}\right)],
\end{split}
\end{equation}
where $\hat{\theta}_1$, $\hat{\theta}_2$, and $\hat{x}_2$ are obtained from the corresponding observer. For the I$\&$I implementation, these signals are generated by \eqref{eqn:adaptive1}-\eqref{eqn:adaptive2}, whereas for the SM implementation they are generated by \eqref{eqn:SMObserverA}--\eqref{eqn:SMObserverB}. The desired trajectory $x_{1d}$ is defined as
\begin{equation}\label{eqn:Qd}
x_{1d}(t)=0.3\left[1-e^{-2.0t^3}\right]\sin(7t),
\end{equation}
and the tracking error is defined as $e_1:=x_1-x_{1d}$. The quantities $k_p>0$ and $k_v>0$ are constant controller gains. Under perfect velocity and parameter estimation, the closed-loop tracking-error dynamics reduce to
\begin{equation}
\ddot{e}_1+k_v\dot{e}_1+k_pe_1=0.
\end{equation}
The behavior of this standard second-order system is well established: its natural frequency and damping ratio are given by $\omega_n=\sqrt{k_p}$ and $\zeta=\frac{k_v}{2\sqrt{k_p}}$, respectively. In \cite{FairComparison}, it is claimed that the closed-loop transient response is governed by the predefined controller gains $k_p$ and $k_v$ under the assumption of perfect estimation. In practice, however, perfect velocity and parameter estimation cannot be ensured. Consequently, the experimental transient response also depends on the observer dynamics and estimation errors, as shown in the following results.
\par
\begin{figure*}[ht!]
  \centering
  \includegraphics[width=\textwidth]{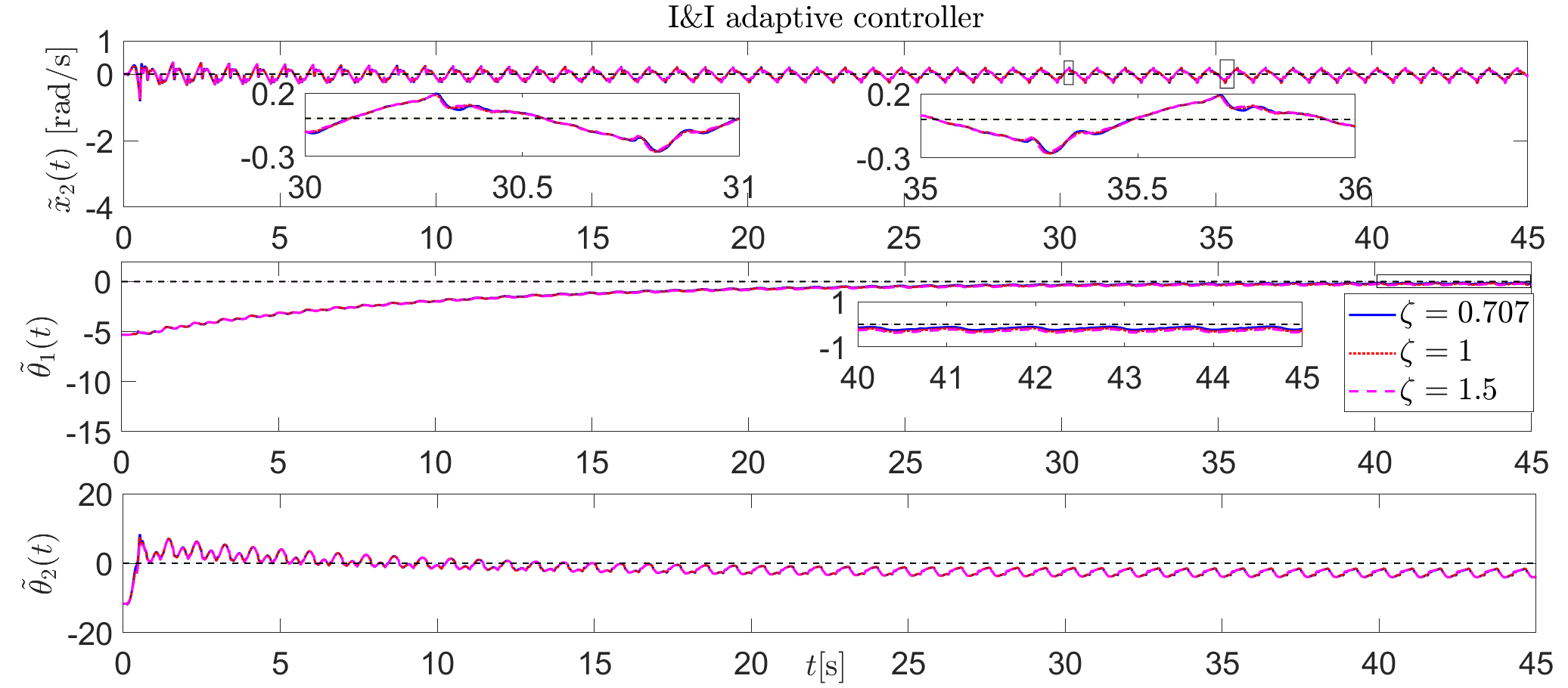}
  \caption{Velocity- and friction-parameter-estimation errors for the I$\&$I implementation using different values of the damping ratio $\zeta$.}
  \label{fig:F4}
\end{figure*}

\begin{figure*}[hb!]
  \centering
  \includegraphics[width=\textwidth]{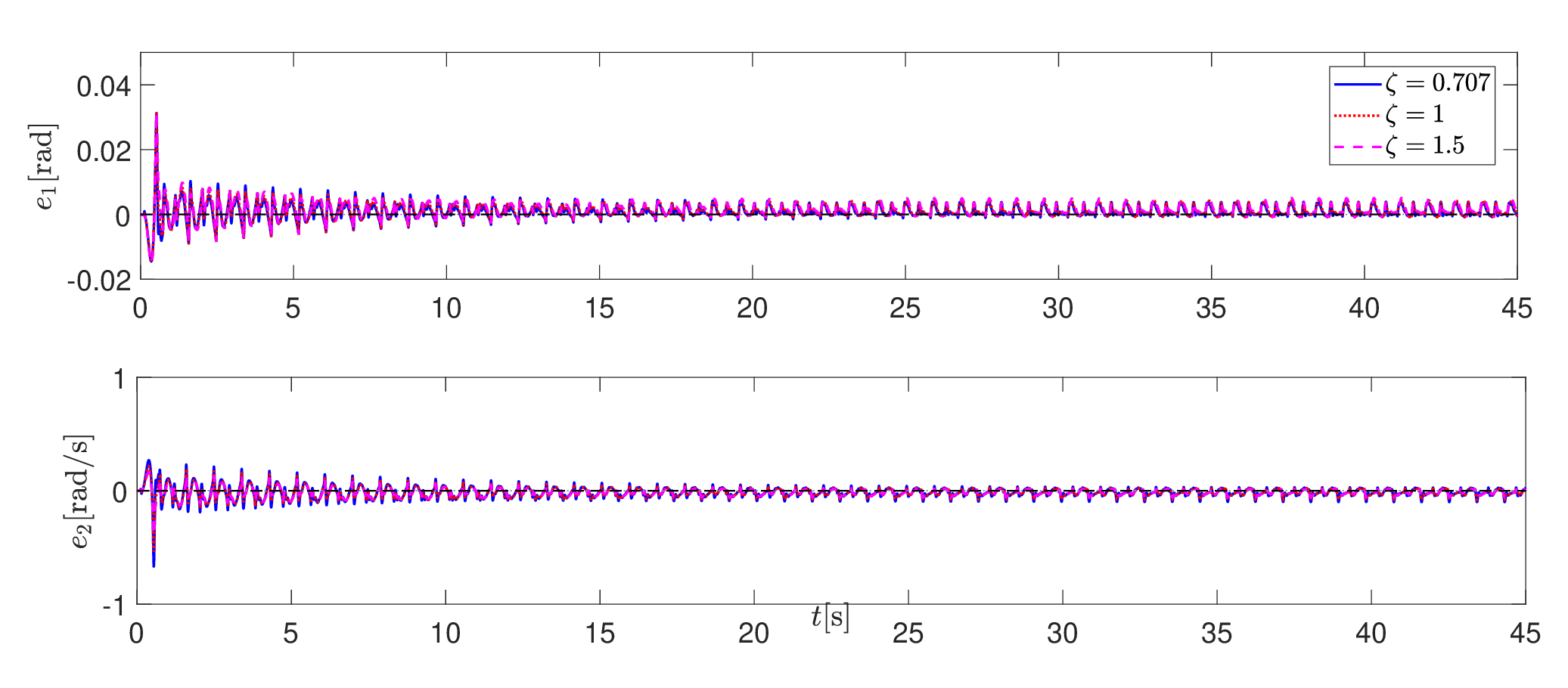}
  \caption{Trajectory-tracking errors for the I$\&$I implementation using different values of the damping ratio $\zeta$.}
  \label{fig:F5}
\end{figure*}

\begin{figure*}[ht!]
  \centering
  \includegraphics[width=\textwidth]{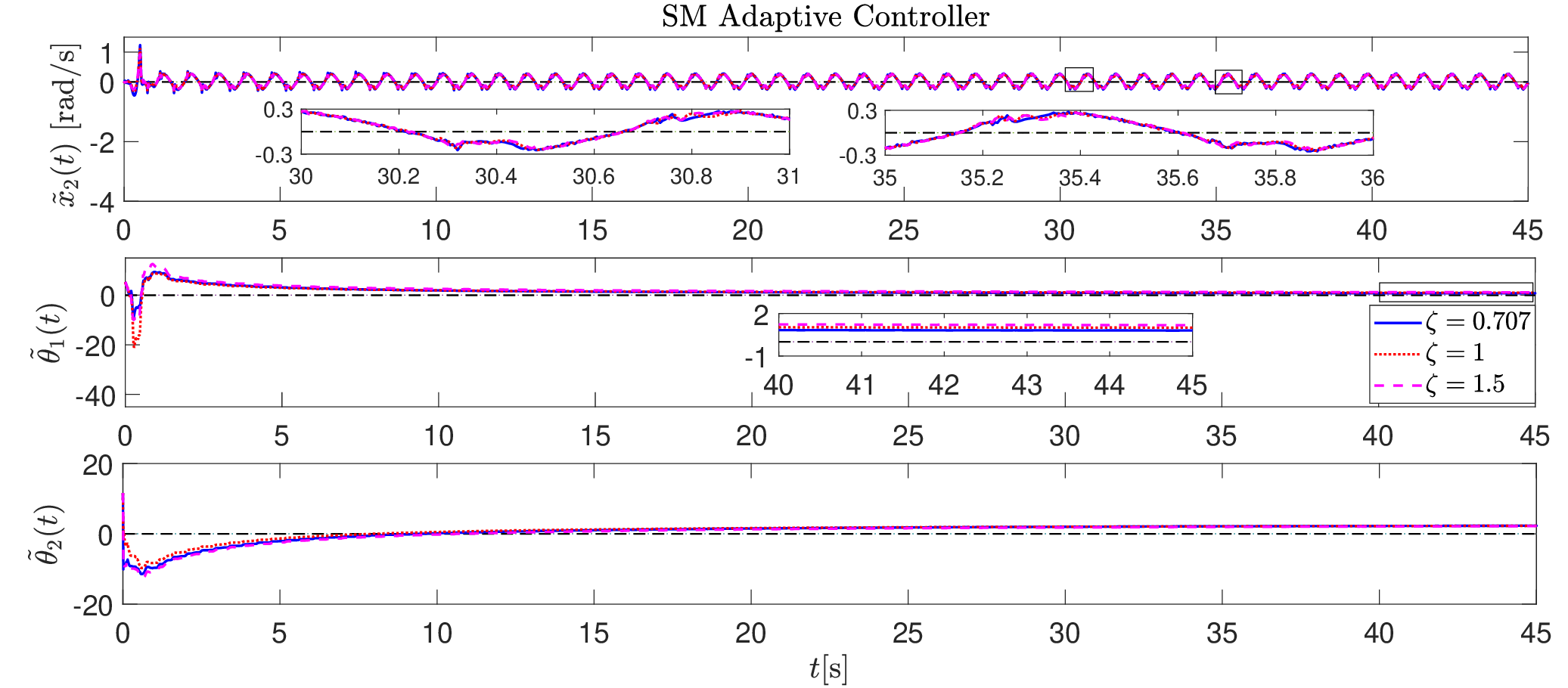}
  \caption{Velocity- and friction-parameter-estimation errors for the SM implementation using different values of the damping ratio $\zeta$.}
  \label{fig:F6}
\end{figure*}

\begin{figure*}[hb!]
  \centering
  \includegraphics[width=\textwidth]{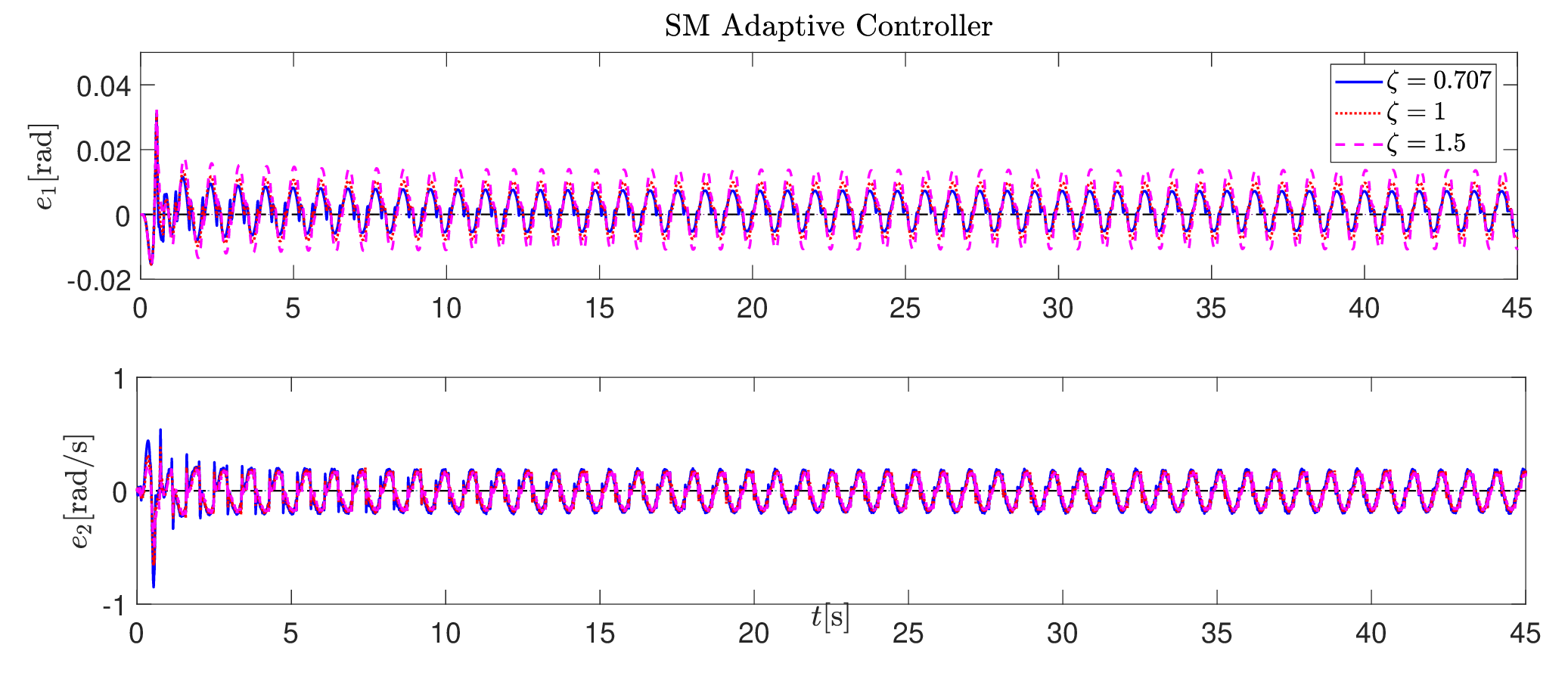}
  \caption{Trajectory-tracking errors for the SM implementation using different values of the damping ratio $\zeta$.}
  \label{fig:F7}
\end{figure*}

 The following conditions are used in all subsequent experiments. Each experiment starts from the initial conditions $q(0)=\dot{q}(0)=0$, and the parameter $\vartheta$ is set to $260$. For the I$\&$I observer/controller, the gains are selected as $k_1=22.5$, $\gamma_1=0.03$, and $\gamma_2=0.65$. The corresponding observer states are initialized as $\theta_{1I}(0)=\theta_{2I}(0)=x_{2I}(0)=0$, which yields $\hat{\theta}_1(0)=\hat{\theta}_2(0)=\hat{x}_2(0)=0$. For the SM observer, the initial conditions are set to $\hat{x}_1(0)=\hat{x}_2(0)=0$ and $\Gamma(0)=100I_2$. The bound parameter is selected as $L=32.4$, although no independent \emph{a priori} information is available to verify that this value satisfies the condition in \eqref{eqn:SMTuningCurrentTest}. Applying the tuning prescription in \eqref{eqn:SMTuningCurrentTest} yields $\alpha_1=1.1L=35.64$ and $\alpha_2=1.5\sqrt{L}\approx8.54$. The normalized friction parameters supplied \emph{a priori} are set to $\frac{\bar{\theta}_1}{J}=\frac{\bar{\theta}_2}{J}=0.01$. For both control implementations, the proportional gain is fixed at $k_p=1600$, which is the same value employed in the original experiment reported in \cite{CervantesPerez2026Experimental}. This selection gives the nominal natural frequency $\omega_n=40~\mathrm{rad/s}$. To generate the three representative types of nominal second-order response, the derivative gain is assigned the values $k_v=56.57$, $k_v=80$, and $k_v=120$. These values correspond to an underdamped response with $\zeta\approx0.707$, a critically damped response with $\zeta=1$, and an overdamped response with $\zeta=1.5$, respectively. These three cases allow us to assess whether the experimental transient response changes as predicted by the nominal characteristic equation. Figures~\ref{fig:F4}--\ref{fig:F7} present the corresponding results.  Figures~\ref{fig:F4} and~\ref{fig:F5} show the closed-loop response of the I$\&$I implementation for the three nominal damping cases. Figure~\ref{fig:F4} presents the observer-related signals and their corresponding estimation errors. Despite the substantial differences among the three nominal damping ratios, the experimental transient responses are practically indistinguishable. Moreover, the velocity- and parameter-estimation errors do not converge exactly to zero during the experimental interval. Therefore, the perfect-estimation assumption required to reduce the closed-loop dynamics to the nominal characteristic equation is not satisfied, and the transient response cannot be determined solely from the prescribed values of $k_p$ and $k_v$. Figure~\ref{fig:F5} shows the position- and velocity-tracking errors, where $e_2=x_2-\dot{x}_{1d}$ and $x_2$ denotes the actual velocity. The close agreement among the three responses confirms that, under the tested conditions, changing the nominal damping ratio does not produce an appreciable change in the experimental transient response. The results also demonstrate satisfactory performance of the I$\&$I implementation in all three cases, with smooth error signals and no visible chattering.
\par Figures~\ref{fig:F6} and~\ref{fig:F7} present the corresponding results for the SM implementation. Once again, varying the nominal damping ratio does not produce an appreciable change in the closed-loop signals. Figure~\ref{fig:F6} shows the velocity- and parameter-estimation errors. \newpage
\clearpage
\clearpage
The parameter-estimation errors approach nonzero constant values, indicating that the estimated friction parameters do not converge to their identified values. Figure~\ref{fig:F7} shows the tracking errors $e_1$ and $e_2$. These results exhibit substantially better tracking performance than that originally reported in \cite{CervantesPerez2026Experimental}, although the error magnitudes remain slightly larger than those obtained with the I$\&$I implementation in Figure~\ref{fig:F5}.
\begin{figure*}[ht!]
  \centering
  \includegraphics[width=\textwidth]{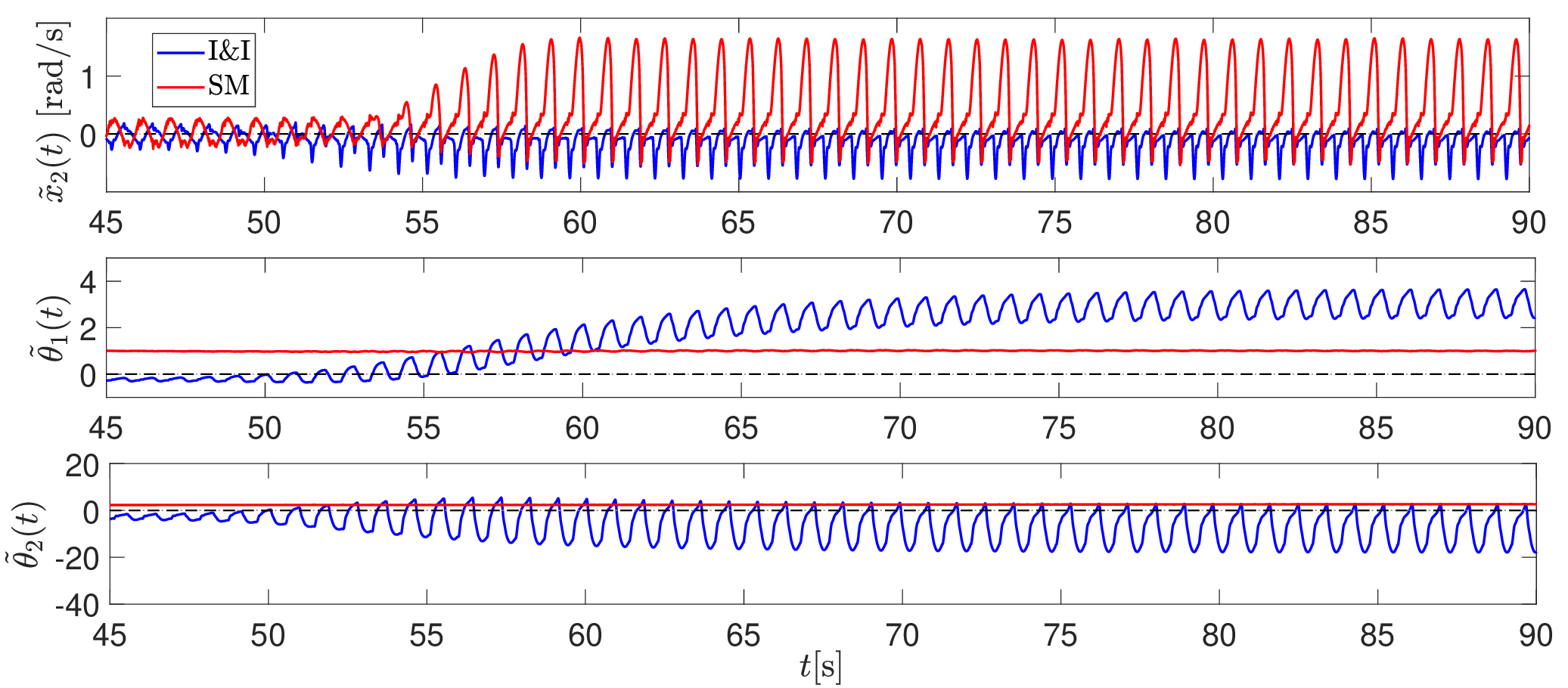}
  \caption{Velocity- and friction-parameter-estimation errors for both controllers following a change in the reference trajectory, with $L=32.4$ used for the SM observer.}
  \label{fig:F8}
\end{figure*}
\par At this point, both control implementations have been evaluated under three different nominal damping conditions. The results show that, under the tested conditions, changing the damping ratio does not appreciably modify the experimental transient response. Therefore, the nominal characteristic equation alone is insufficient to predict the actual closed-loop behavior when nonzero observer and parameter-estimation errors are present.
\par
\begin{figure*}[h!]
  \centering
  \includegraphics[width=\textwidth]{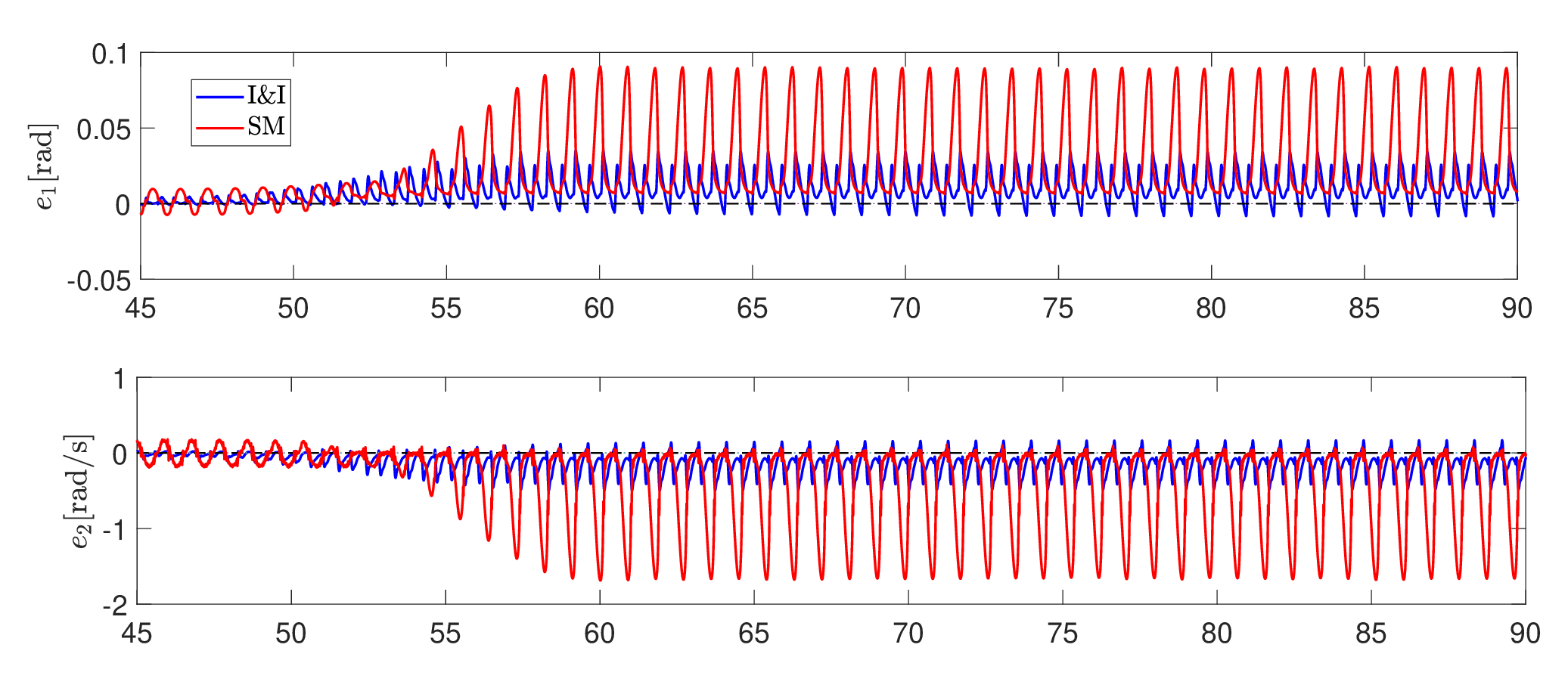}
  \caption{Trajectory-tracking errors for both controllers following a change in the reference trajectory, with $L=32.4$ used for the SM observer.}
  \label{fig:F9}
\end{figure*}

\begin{figure*}[h!]
  \centering
  \includegraphics[width=\textwidth]{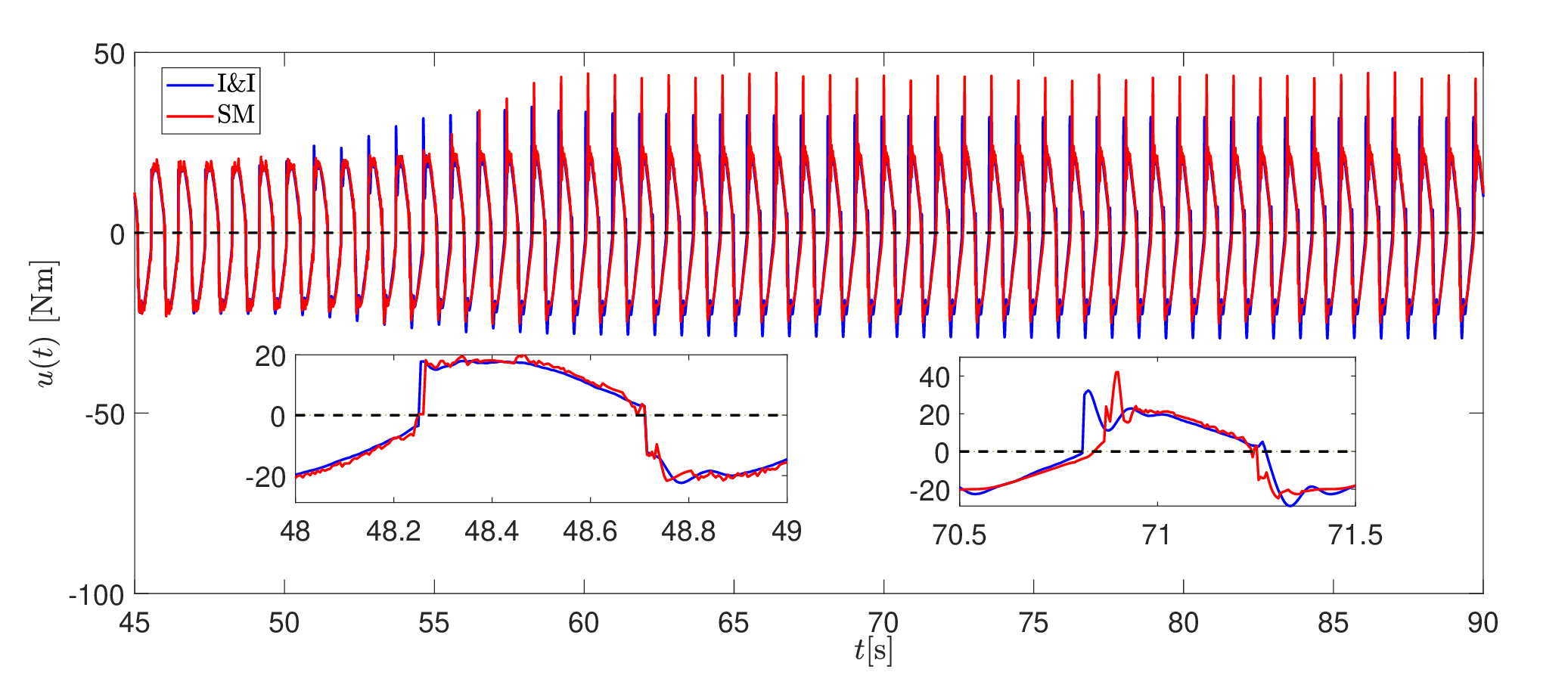}
  \caption{Control-input behavior following a change in the reference trajectory for $L=32.4$.}
  \label{fig:F10}
\end{figure*}

\begin{figure*}[h!]
  \centering
  \includegraphics[width=\textwidth]{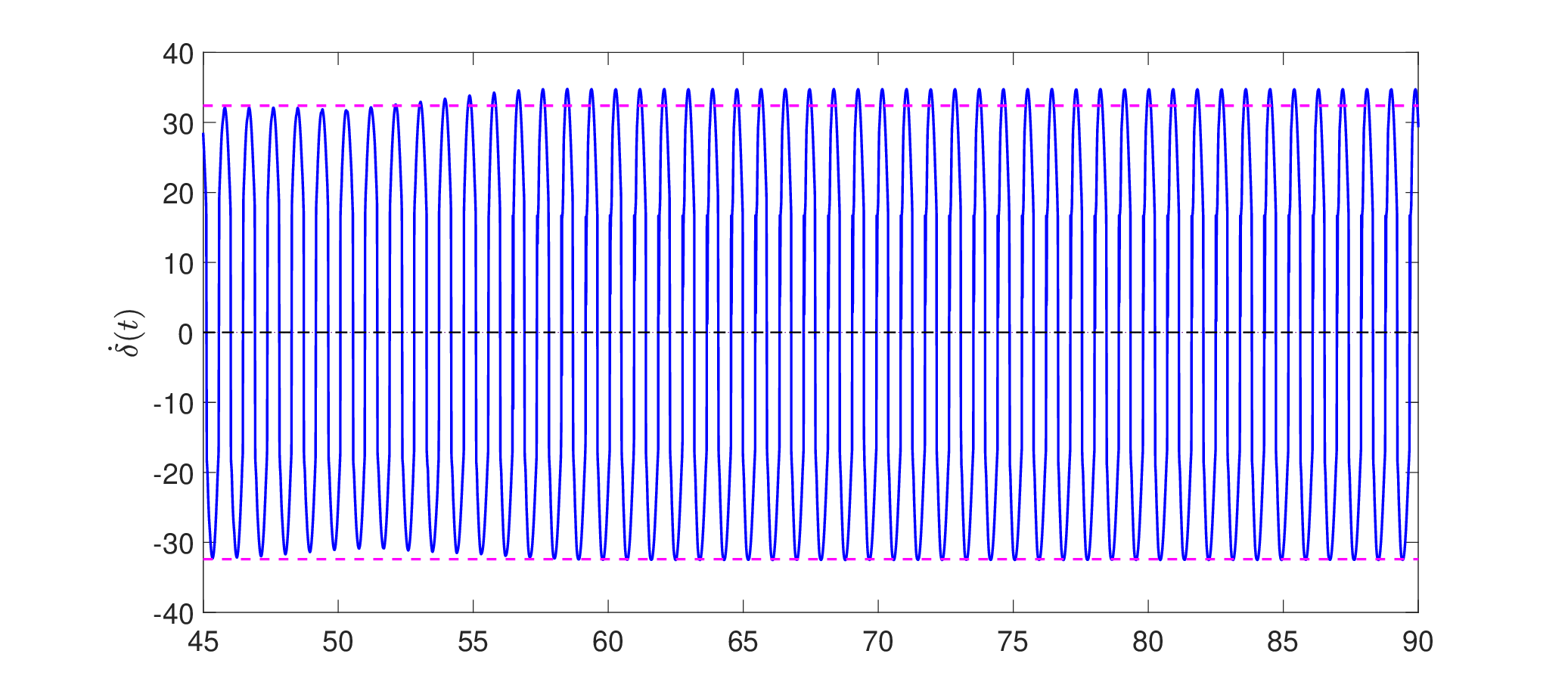}
  \caption{Behavior of the reconstructed perturbation $\dot{\delta}(t)$ following a change in the reference trajectory for $L=32.4$.}
  \label{fig:F11}
\end{figure*}

 We next select the critically damped configuration, $\zeta=1$, for both controllers and evaluate their responses following a change in the desired trajectory. This case corresponds to $k_p=1600$ and $k_v=80$. The additional test provides an experimental assessment of robustness with respect to variations in the reference trajectory. During the first $45~\mathrm{[s]}$, the desired trajectory is the same as that defined in \eqref{eqn:Qd}. It is subsequently blended over a $15~\mathrm{[s]}$ interval with a second sinusoidal trajectory having an offset of $\frac{\pi}{4}~\mathrm{rad}$ and an amplitude of $0.35~\mathrm{[rad]}$. Define
\begin{align}
r_1(t)=&0.3\left(1-e^{-2t^3}\right)\sin(7t),\\
r_2(t)=&\frac{\pi}{4}+0.35\sin\left[7(t-45)\right],\\
s(t)=&\frac{t-45}{15},\\
\beta(s)=&10s^3-15s^4+6s^5.
\end{align}
The modified desired trajectory is then given by
\begin{equation}
\label{eqn:ModifiedDesiredTrajectory}
x_{1d}(t)=
\begin{cases}
r_1(t), & 0\leq t\leq45,\\[3pt]
\begin{aligned}
&[1-\beta(s(t))]r_1(t)\\
&\quad+\beta(s(t))r_2(t),
\end{aligned}
&45<t<60,\\[3pt]
r_2(t), &t\geq60.
\end{cases}
\end{equation}
The quintic blending function satisfies $\beta(0)=0$ and $\beta(1)=1$, while its first and second derivatives vanish at both endpoints. Consequently, the desired position, velocity, and acceleration remain continuous at $t=45~\mathrm{[s]}$ and $t=60~\mathrm{[s]}$. This construction is proposed to ensure a \emph{smooth} transition between the two reference trajectories. After the transition, the reference oscillates around the constant offset of $\frac{\pi}{4}~\mathrm{[rad]}$ with an amplitude of $0.35~\mathrm{[rad]}$ and an angular frequency of $7~\mathrm{[rad/s]}$.  The same observer and controller gains used in the preceding experiment are retained, and the results are presented in Figures~\ref{fig:F8}--\ref{fig:F11}. Only the interval from $t=45~\mathrm{[s]}$ onward is shown because the behavior during the preceding interval has already been presented in Figures~\ref{fig:F4}--\ref{fig:F7}. Figure~\ref{fig:F8} shows the velocity- and parameter-estimation errors for both controllers. A deterioration in estimation performance is observed for both implementations after the reference transition begins. This deterioration is considerably more pronounced for the SM observer, whose velocity-estimation error develops a clearly visible nonzero offset. The change in the reference trajectory also modifies the excitation signals experienced by the adaptive observers, thereby altering the evolution of the parameter-estimation errors. Although the I$\&$I observer is also affected by the reference change, it maintains better overall estimation performance. Figure~\ref{fig:F9} shows the trajectory-tracking errors $e_1$ and $e_2$ for both controllers. A deterioration in tracking performance is again observed for both implementations after the reference transition. This deterioration is more pronounced for the SM controller, whose tracking-error signals exhibit a clearly visible unintended offset.
Figure~\ref{fig:F10} shows the corresponding control signals. Chattering becomes apparent in the SM control input, whereas the I$\&$I implementation produces a smooth control signal. At this stage, the overall experimental comparison favors the I$\&$I controller. Nevertheless, to address the possibility that the observed SM performance results from inadequate observer tuning, we retain the I$\&$I configuration as a fixed benchmark and attempt to improve the SM response by following the tuning prescription in \eqref{eqn:SMTuningCurrentTest}.
\begin{figure*}[h!]
  \centering
  \includegraphics[width=\textwidth]{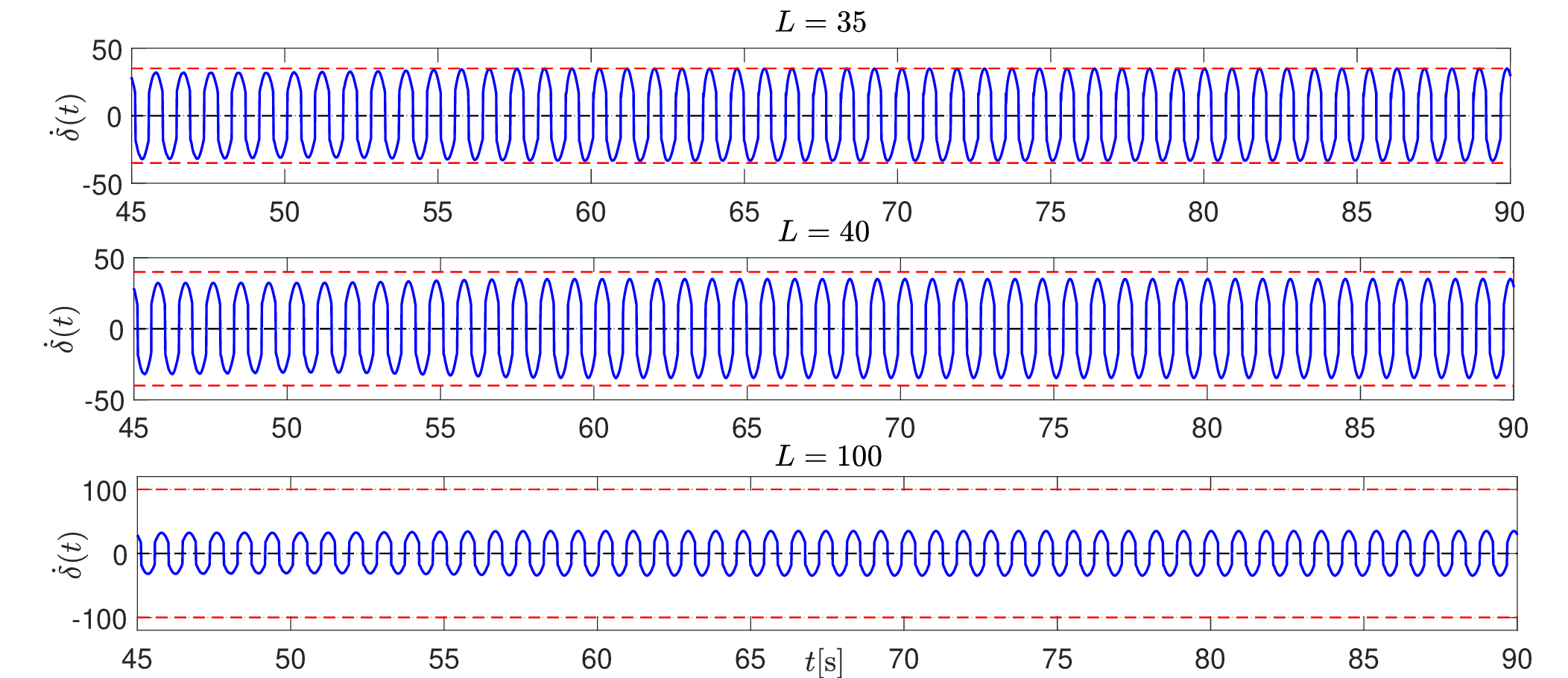}
  \caption{Behavior of the reconstructed \emph{perturbation} $\dot{\delta}(t)$ relative to its corresponding prescribed bound $L$ following a change in the reference trajectory, with the SM observer evaluated using different values of $L$.}
  \label{fig:F12}
\end{figure*} 
\begin{figure*}[h!]
  \centering
  \includegraphics[width=\textwidth]{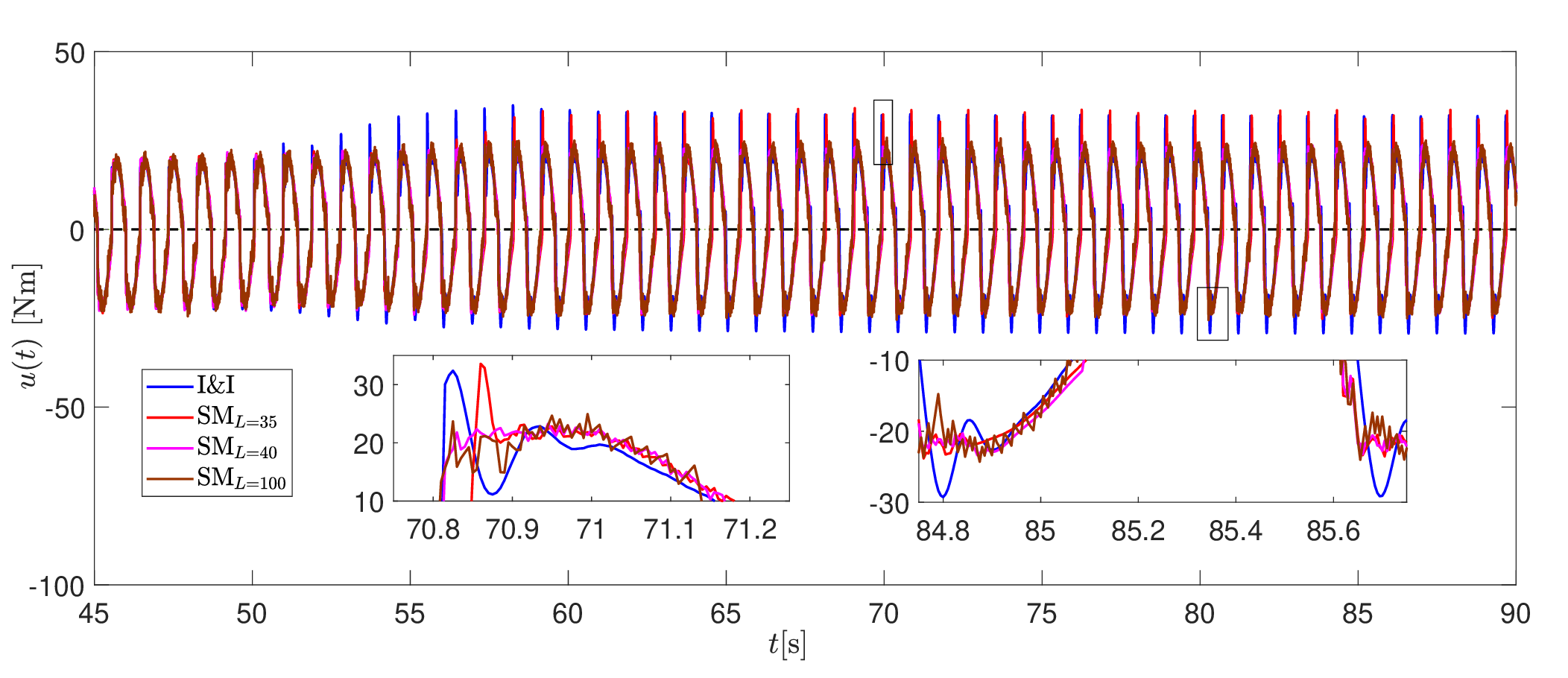}
  \caption{Control signals for the I$\&$I and SM controller configurations following a change in the reference trajectory, with the SM controller evaluated using different values of $L$.}
  \label{fig:F15}
\end{figure*}
The preceding experiments showed that varying the nominal damping ratio did not produce an appreciable change in the transient response. Therefore, with the controller gains and the \emph{a priori} friction parameters held fixed, the remaining tuning variable in the present protocol is $L$, together with the observer gains determined from it. 
\begin{figure*}[h!]
  \centering
  \includegraphics[width=\textwidth]{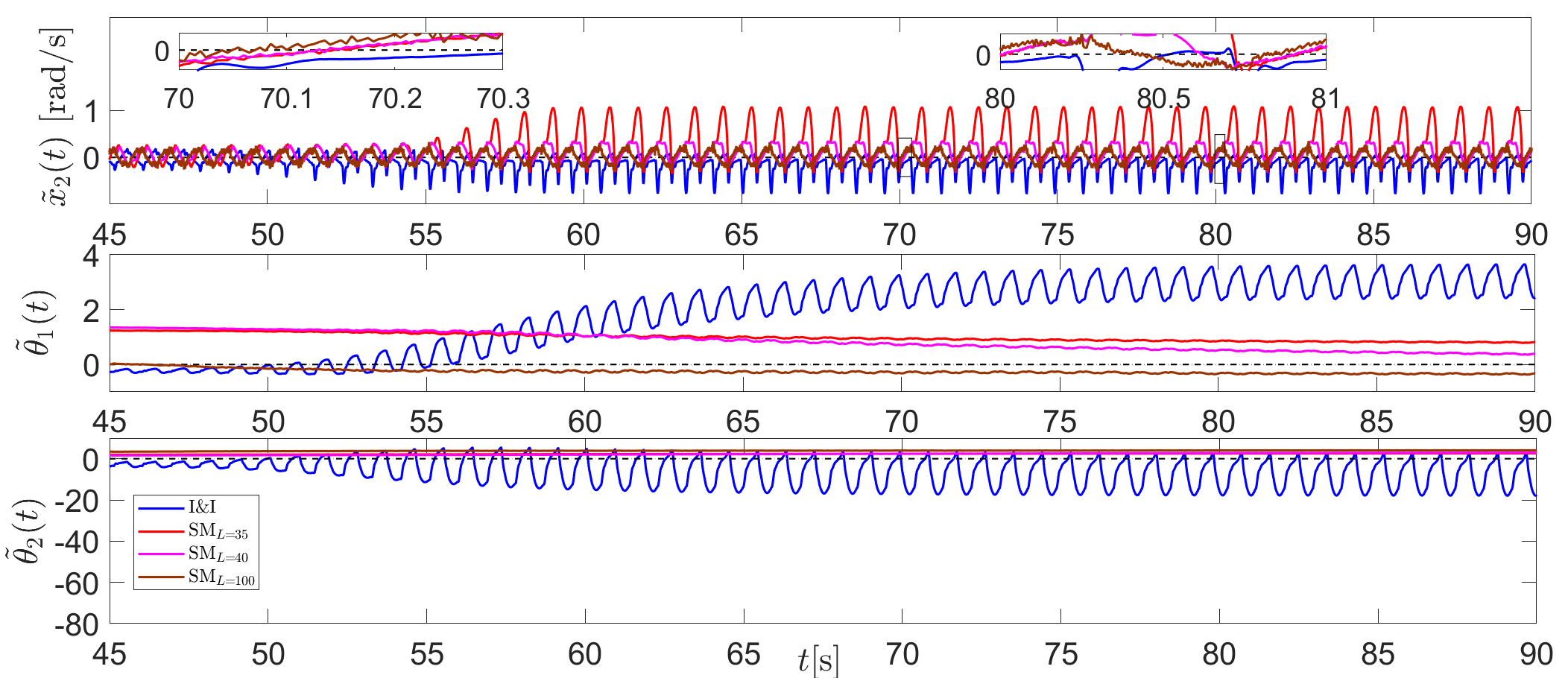}
  \caption{Velocity- and friction-parameter-estimation errors for the I$\&$I and SM implementations following a change in the reference trajectory, with the SM observer evaluated using different values of $L$.}
  \label{fig:F13}
\end{figure*}
\begin{figure*}[h!]
  \centering
  \includegraphics[width=\textwidth]{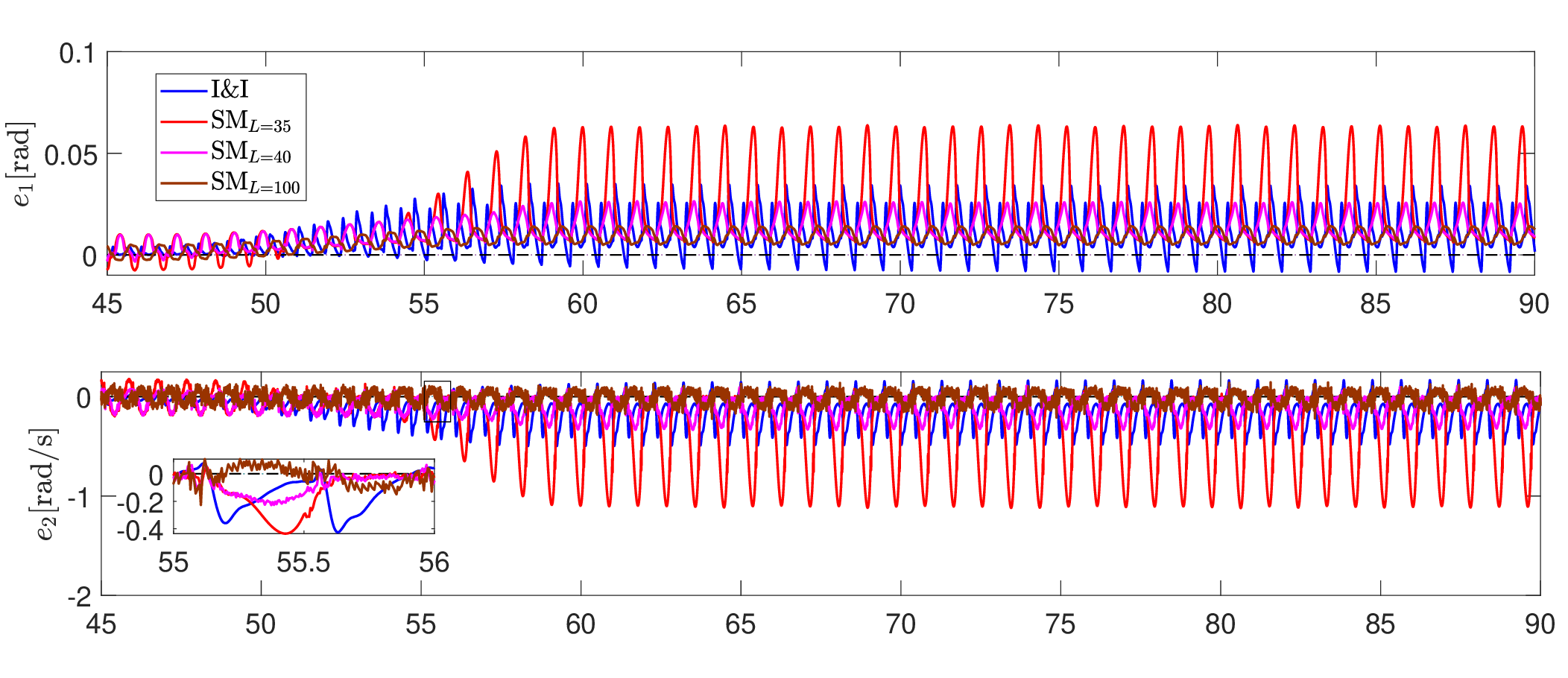}
  \caption{Trajectory-tracking error signals for the I$\&$I and SM controller configurations following a change in the reference trajectory, with the SM controller evaluated using different values of $L$.}
  \label{fig:F14}
\end{figure*}
Recall that $L=32.4$ was initially selected without an independent \emph{a priori} procedure for verifying the inequality in \eqref{eqn:SMTuningCurrentTest}. Figure~\ref{fig:F11} shows the reconstructed perturbation $\dot{\delta}(t)$ during and after the smooth reference transition, while the dotted horizontal line represents the selected bound $L=32.4$. Although this value happens to dominate the reconstructed perturbation during the initial operating interval, $\dot{\delta}(t)$ exceeds $L$ after the reference transition. Consequently, the sufficient condition required by \eqref{eqn:SMTuningCurrentTest} is no longer satisfied during that portion of the experiment. This exceedance does not, by itself, prove that violating the bound causes the observed deterioration, because the condition is sufficient rather than necessary. Nevertheless, it shows that the theoretical tuning hypothesis is no longer verified and coincides with the degradation observed in the estimation and tracking signals. More importantly, the adequacy of the initially selected value of $L$ can only be determined \emph{a posteriori}, after conducting the experiment and reconstructing $\dot{\delta}$. Following the logic of the proposed tuning prescription, a natural remedy is to increase $L$ and, consequently, the observer gains. We therefore repeat the experiment using $L=35$, $L=40$, and $L=100$. The corresponding gain pairs $(\alpha_1,\alpha_2)$ are $(38.5,8.87)$, $(44,9.49)$, and $(110,15)$, respectively. Each pair is obtained directly from \eqref{eqn:SMTuningCurrentTest}. These selections satisfy the prescribed algebraic gain relationships by construction; however, whether each selected value of $L$ actually bounds $\dot{\delta}(t)$ must still be verified separately for the resulting experimental trajectory. Figures~\ref{fig:F12}--\ref{fig:F15} show the results of the retuning experiments. First, Figure~\ref{fig:F12} compares the reconstructed perturbation $\dot{\delta}(t)$ with the corresponding selected bound $L$ for each SM configuration. Changing $L$ modifies the observer gains $\alpha_1$ and $\alpha_2$ and, consequently, the observer-state trajectories. Because $\hat{x}_2$ appears explicitly in the reconstruction of $\dot{\delta}$, the reconstructed perturbation also changes with the selected value of $L$. Therefore, the quantity whose bound is used to tune the observer is not independent of the resulting observer gains. This dependence illustrates the experiment-dependent circularity that arises when $L$ is selected or adjusted \emph{a posteriori} from the reconstructed perturbation. For all the retuned configurations, the selected value of $L$ bounds $\dot{\delta}(t)$ throughout the recorded experimental interval. As expected, increasing $L$ produces a larger empirical margin between the reconstructed perturbation and its selected bound. Nevertheless, this observation verifies the inequality only for each recorded trajectory and does not establish a uniform bound for different references, initial conditions, disturbances, or operating conditions.

Figure~\ref{fig:F13} shows the velocity- and parameter-estimation error signals. Chattering is visible in the velocity-estimation error for all three selected values of $L$ and becomes increasingly pronounced as $L$ is increased. The largest value of $L$ substantially reduces and nearly eliminates the nonzero offset; however, this improvement is obtained at the cost of considerably greater chattering. Therefore, simply increasing $L$ does not produce an unequivocal improvement in the overall estimation performance.

Figure~\ref{fig:F14} shows the trajectory-tracking errors. As $L$ is increased, the nonzero offset is progressively reduced and is nearly eliminated for the largest value of $L$. However, chattering becomes increasingly evident in the tracking-error signals.
Finally, Figure~\ref{fig:F15} shows the corresponding control signals, which clearly reflect the chattering observed in the SM closed-loop responses. As reported in \cite{CervantesPerez2026Experimental}, this phenomenon is not limited to the oscillations visible in the recorded signals; it is also manifested as audible high-frequency noise. Video and audio evidence of the experimental behavior can be consulted at \href{https://youtu.be/5mFO8tQcrf0}{https://youtu.be/5mFO8tQcrf0}.
\section{Discussion and concluding remarks}
This technical report has presented several open-loop and closed-loop tests conducted on the same direct-drive pendular platform to evaluate the I$\&$I and the SM adaptive observers implementations under the tuning conditions proposed in \cite{FairComparison}.

In particular, every SM observer gain pair was selected to satisfy the $L$-free relationship in \eqref{eqn:LFreeRelationshipCurrent}. The experiments separately varied the excitation signal, the fixed \emph{a priori} friction-parameter values, the nominal closed-loop pole placement, and the SM observer gains. The open-loop experiments demonstrate that the \emph{a priori} friction information has a significant influence on the observer response. Moreover, the trajectory-tracking experiments demonstrate that, even when the tuning rules of \cite{FairComparison} are followed, chattering remains present. Although it is attenuated in certain scenarios, it is not eliminated.
\par We conclude this report with the following comments regarding the response presented in \cite{FairComparison}.
\begin{enumerate}[label={[\arabic*]}]
  \item Although the tuning guideline proposed in \cite{FairComparison} is presented as \emph{fully transparent}, it remains unclear what $L$ represents, how $\dot{\delta}$ is defined, and under which conditions the strong inequality $L>\sup|\dot{\delta}|$ can be guaranteed. Therefore, the proposed rule still depends on unspecified quantities and conditions. The $L$-free relationship used in this report eliminates $L$ from the observer-gain ratio; however, the upper-bound condition remains part of the stability analysis. Thus, it is unclear how this assumption can be verified in practical applications.
\item The experiments in \cite{FairComparison} also use a different type of actuator. Their platform employs a DC motor driven through PWM, whereas our experimental device uses a direct-drive motor operated in torque mode. This distinction is important because the controller design assumes that the control input is the applied torque, whereas the physical input to their system appears to be either the motor voltage or the PWM duty cycle. The conversion between these quantities and the effect of the actuator dynamics are not clearly explained. Furthermore, the possible influence of the PWM frequency and sampling period on the observed chattering is not analyzed, even though these implementation factors could contribute to its attenuation or modify its experimental manifestation.
\item Describing our experiments as ``rigged'' introduces a judgment that is not supported by a technical argument. Such terminology does not contribute to an objective comparison and introduces precisely the type of unsupported assessment criticized in the discussion of our original article. Experimental conclusions should instead be based on matched conditions, clearly defined tuning procedures, and measurable performance indicators.
\item An additional issue not considered in \cite{FairComparison} is that the I$\&$I scheme of \cite{romero2025adaptive} and the adaptive SM of \cite{SM1} were designed to estimate unknown friction parameters. Their objective is not limited to reconstructing the unmeasured velocity or minimizing the tracking error; under the corresponding excitation conditions, they also seek convergence of the friction-parameter estimates to their actual values. This objective is demonstrated in the trajectory-tracking experiments, where the parameter-estimation errors approach zero. Therefore, a comparison focused exclusively on minimizing the tracking error overlooks one of the principal objectives of the original adaptive designs.
\end{enumerate}
\par The experiments presented in this report show that following the tuning rules proposed in \cite{FairComparison} can improve the transient response in certain operating conditions. Nevertheless, chattering remains visible in the velocity estimates, tracking errors, and control inputs. Therefore, satisfying the proposed tuning relationship is not sufficient to guarantee a chattering-free experimental implementation. This conclusion is consistent with the caution expressed by Utkin et al., who observed that optimistic perspectives concerning chattering suppression in higher-order SM ``do not look clear for us and need more convincing arguments'' \cite[p.~10259]{UtkinEtAl2020}.


\end{document}